\documentclass[sigconf]{acmart}
\AtBeginDocument{%
  \providecommand\BibTeX{{%
    Bib\TeX}}}

\copyrightyear{2026}
\acmYear{2026}
\setcopyright{cc}
\setcctype{by}
\acmConference[FORGE '26]{2026 IEEE/ACM Third International Conference on AI Foundation Models and Software Engineering}{April 12--13, 2026}{Rio de Janeiro, Brazil}
\acmBooktitle{2026 IEEE/ACM Third International Conference on AI Foundation Models and Software Engineering (FORGE '26), April 12--13, 2026, Rio de Janeiro, Brazil}
\acmPrice{}
\acmDOI{10.1145/3793655.3793737}
\acmISBN{979-8-4007-2477-0/2026/04}

\usepackage{amsmath,amsfonts}
\usepackage{multirow}
\usepackage{algorithmic} 

\usepackage{graphicx}
\usepackage{textcomp}
\usepackage{xcolor}
\usepackage{booktabs}
\usepackage{adjustbox}
\usepackage{subcaption}
\usepackage{float}
\usepackage{graphicx}
\usepackage{caption}
\usepackage{subcaption}
\usepackage[ruled, vlined, linesnumbered]{algorithm2e}
\SetAlFnt{\small}
\SetAlCapNameFnt{\small}
\SetAlCapFnt{\small}
\SetFuncSty{textsc}
\AtBeginEnvironment{algorithm}{
	\DontPrintSemicolon
}
\SetKwRepeat{Do}{do}{while}

\newsavebox\CBox

\usepackage{hyperref}

\def\BibTeX{{\rm B\kern-.05em{\sc i\kern-.025em b}\kern-.08em
    T\kern-.1667em\lower.7ex\hbox{E}\kern-.125emX}}

\let\oldnl\nl% Store \nl in \oldnl
\newcommand{\nonl}{\renewcommand{\nl}{\let\nl\oldnl}}% Remove line number for one line
\usepackage{tcolorbox}
\newtcolorbox{custombox}[1]{
	colback=gray!10,
	colframe=gray!20,
	left=1mm,
	right=1mm,
	top=1mm,
	bottom=1mm,
	fonttitle=\bfseries,
	arc=2mm,
	leftrule=0mm,
	rightrule=.5mm,
	toprule=0mm,
	bottomrule=.5mm,
	notitle,
	before=\par\smallskip\noindent,
	before upper={\textbf{#1: } },
}

\begin{document}

\title{Exploring the Potential of Large Language Models in Simulink-Stateflow Mutant Generation }

\author{Pablo Valle}
\affiliation{%
  \institution{Mondragon University}
  \city{Mondragon}
  \state{Gipuzkoa}
  \country{Spain}}
\email{pvalle@mondragon.edu}

\author{Shaukat Ali}
\affiliation{%
  \institution{Simula Research Laboratory and}
  \institution{Oslo Metropolitan University}
  \city{Oslo}
  \country{Norway}}
\email{shaukat@simula.no}
\email{shaukata@oslomet.no}

\author{Aitor Arrieta}
\affiliation{%
  \institution{Mondragon University}
  \city{Mondragon}
  \state{Gipuzkoa}
  \country{Spain}}
\email{aarrieta@mondragon.edu}

\begin{abstract}
Mutation analysis is a powerful technique for assessing test-suite adequacy, yet conventional approaches suffer from generating redundant, equivalent, or non-executable mutants. These challenges are particularly amplified in Simulink-Stateflow models due to the  hierarchical structure these models have, which integrate continuous dynamics with discrete-event behaviors and are widely deployed in safety-critical Cyber-Physical Systems (CPSs). While prior work has explored machine learning and manually engineered mutation operators, these approaches remain constrained by limited training data and scalability issues. Motivated by recent advances in Large Language Models (LLMs), we investigate their potential to generate high-quality, domain-specific mutants for Simulink-Stateflow models. We develop an automated pipeline that converts Simulink-Stateflow models to structured JSON representations and systematically evaluates different mutation and prompting strategies across eight state-of-the-art LLMs. Through a comprehensive empirical study involving 38,400 LLM-generated mutants across four Simulink-Stateflow models, we demonstrate that LLMs generate mutants up to 13x faster than a manually engineered mutation-based baseline while producing significantly fewer equivalent and duplicate mutants and consistently achieving superior mutant quality. Moreover, our analysis reveals that few-shot prompting combined with low-to-medium temperature values yields optimal results. We provide an open-source prototype tool and release our complete dataset to facilitate reproducibility and advance future research in this domain.
\end{abstract}

\begin{CCSXML}
<ccs2012>
   <concept>
       <concept_id>10011007.10011074.10011099</concept_id>
       <concept_desc>Software and its engineering~Software verification and validation</concept_desc>
       <concept_significance>500</concept_significance>
       </concept>
   <concept>
        <concept_id>10011007.10011074.10011099.10011102.10011103</concept_id>
       <concept_desc>Software and its engineering~Software testing and debugging</concept_desc>
       <concept_significance>500</concept_significance>
       </concept>
   <concept>
       <concept_id>10011007.10011074.10011092.10011094</concept_id>
       <concept_desc>Software and its engineering~Flowcharts</concept_desc>
       <concept_significance>500</concept_significance>
       </concept>

 </ccs2012>
\end{CCSXML}

\ccsdesc[500]{Software and its engineering~Software verification and validation}
\ccsdesc[500]{Software and its engineering~Software testing and debugging}
\ccsdesc[500]{Software and its engineering~Flowcharts}

\keywords{Large Language Models, Cyber-Physical Systems, Mutation Analysis, Simulink, Stateflow Diagrams}

\maketitle
\vspace{-0.1cm}
\section{Introduction}

Mutation analysis~\cite{budd1980mutation} is an effective software testing method used to assess the adequacy of test suites by inserting artificial faults into a program's source code using mutation operators. These operators imitate developers' common mistakes, producing modified program versions, known as mutants. The effectiveness of a test suite is then measured by its ability to detect behavioral differences between the mutants and the original program. A mutant is considered \textit{detected} if at least one test case produces a different output than the original model; otherwise, it is considered \textit{alive}. The mutation score, i.e., the percentage of mutants detected out of the total number of mutants, is the metric that determines the test adequacy. Despite its advantages, mutation testing has drawbacks, such as the generation of duplicated and equivalent mutants, high computational costs, and scalability issues. 

Conventional mutation analysistesting methods rely on rule-based mutation operators~\cite{jia2010analysis, just2014major}, which make simple syntactic changes throughout the code at all possible locations. While effective at producing a large number of mutants, many are redundant or equivalent (i.e., mutants that exhibit no behavioral difference or are semantically identical to the original code), offering no additional value yet requiring increasing computational cost. To address this issue, recent studies~\cite{tufano2019learning, patra2021semantic, tian2022learning} have explored using machine learning to generate mutants from historical project data and identified patterns. However, these techniques remain limited by the availability and quality of training data~\cite{ojdanic2023comparing},  often leading to unrealistic or syntactically incorrect mutants that fail to accurately depict real-world faults. An alternative to this is to manually create mutants, but it is expensive and requires substantial domain expertise. Consequently, recent efforts have resorted to usingthe use of Large Language Models (LLMs) for mutant generation~\cite{tip2025llmorpheus, degiovanni2022mu}, leveraging their extensive training on large-scale code data and their ability to reason over semantic and structural patterns. This enables LLMs to produce mutants that go beyond simple syntactic modifications, potentially generating more realistic and diverse mutants. Nevertheless, the effectiveness of LLM-generated mutants remains underexplored and requires explicit assessment.

While the aforementioned challenges are well recognized in source-code mutation, they become even more pronounced in the context of Simulink-Stateflow models \cite{stateflowMATLAB}, which are widely used in developing embedded systems. Unlike traditional source code, Simulink-Stateflow models integrate continuous-time dynamics with discrete-event behaviors, requiring specialized mutation strategies to maintain syntactic validity, adhere to modeling standards, and reflect real-world faults scenarios. Simply altering syntax is often insufficient, as it may produce mutants that either violate model semantics or fail todo not mimic faults made by developers make in practice. Generating meaningful mutants, i.e., mutants that are syntactically valid, semantically consistent, and representative of realistic faults that could affect system behavior, is particularly difficult due to the unique semantics and hierarchical organization of Simulink-Stateflow models, including the connections between states, transitions, and actions triggered by events. Furthermore, as with source-code mutation, it is common to generate redundant and duplicated mutants~\cite{fernandes2017avoiding}, which increases the computational cost of current approaches due to the long execution times of Simulink-Stateflow models. Therefore, it is crucial to develop mutation testing approaches that consider the domain-specific attributes of Simulink-Stateflow models to produce high-quality, meaningful mutants. 

The recent adoption of LLMs in software engineering tasks~\cite{schafer2023empirical, liu2023your, jiang2024survey, liu2024empirical, nejjar2025llms} presents an opportunity to address these domain-specific challenges. Trained on vast software code, LLMs have demonstrated strong capabilities in code generation~\cite{jiang2024survey} and in understanding hierarchical and semantic structures of complex code~\cite{liu2024empirical}. We therefore conjecture that LLMs can be leveraged to generate high-quality, domain-specific mutants for Simulink-Stateflow models, where traditional mutation techniques struggle. To this end, we investigate the potential of LLMs to produce realistic, diverse, and domain-specific mutants for Simulink-Stateflow models.

In summary, our paper makes the following contributions:

% We evaluate our approach on a total of 9 well-known LLMs. Our results provide the following key findings. First, GPT3.5 is more accurate test oracle than other more advanced LLMs (e.g., GPT4, GPT4-o) as well as LLMs that are specifically designed for acting as test oracle of LLM safety (i.e., LlamaGuard). Secondly, our tool is more effective than static baselines. Third, the Llama family of LLMs are generally safer than other LLMs. Fourth, the style and safety category has a huge impact on making the LLM answer unsafe questions. For instance,  \aitor{Give examples} \miriam{the \textit{role play} style, significantly increased the number of unsafe responses in GPT3.5, Misral, and Vicuna, obtaining 153, 60, and 80 unsafe results out of 1260}.

\begin{itemize}
    \item We extensively evaluate eight well-known LLMs, assessing their performance in generating Simulink-Stateflow mutants. WIn our evaluation, we investigate how different mutation strategies, prompting strategies, and temperature settings affect the performance of LLMs.Our findings show that LLMs generate mutants up to 13 times faster than the baseline, with higher quality, greater diversity, and fewer duplicates or equivalents.
    \item We compare different mutation and prompting strategies, and we observe that Global mutation with few-shot prompting achieves the highest generability and compilability, while Local mutation enhances mutant diversity and quality. Notably, the optimal strategy varies across LLMs.
    \item We analyze the non-compilable and non-generable mutations and identify four primary causes of failure: 1) invalid references, 2) incorrect transition mutations, 3) undefined variables, and 4) syntax errors, highlighting the need for structural and syntactic validation in LLM-based mutant generation.
    \item We prototype the method in a tool and provide it along with the generated mutant dataset as open-source on GitHub~\cite{StateflowMutantGeneration2025}. %We will make the replication package~\cite{StateflowMutantGeneration2025} publicly available upon acceptance. 
    
\end{itemize}

The rest of this paper is organized as follows. Section \ref{sec:background} provides the necessary background and foundational concepts relevant to this work, while Section \ref{sec:Approach} describes the proposed approach in detail. The empirical evaluation is presented in Section \ref{sec:Evaluation}, followed by an analysis and discussion of the results in Section \ref{sec:Discussion}. Potential threats to the validity of our study are examined in Section \ref{sec:threats}. Section \ref{sec:RelatedWork} situates this work within the context of existing state-of-the-art research. Finally, Section \ref{sec:Conclusions} concludes the paper and outlines directions for future work.

\section{Background}
\label{sec:background}
\subsection{Stateflow Models}
Stateflow \cite{stateflowMATLAB}, from MathWorks, a modeling notation inside Simulink, is a popular tool for developing complex control systems and state-driven behaviors, especially in CPSs. It combines state machines and flowcharts to illustrate dynamic system behavior, providing an organized structure for representing the systems. A Stateflow model consists of three main components: (1) States, illustrated as rectangular boxes, that stand for particular modes or conditions of the system along with the related behaviors or actions; (2) Transitions, displayed as arrows that connect the states. Transitions establish the conditions, a set of Boolean conditions or events, that trigger a move from one state to another; and (3) Junctions, depicted as small circles, used to merge or split multiple transition paths for complex control flows.
% Stateflow \cite{stateflowMATLAB} is a visual programming language and simulation environment for Simulink created by MathWorks. It is primarily used for modeling and simulating complex control systems and state-based behaviors in various applications. The main components of a Stateflow diagram are the states and the transitions. The states are a set of events shown as a box representing a set of behaviors or actions that the system can perform at that moment. To switch from one state to another, a transition must be triggered. The transitions are represented as arrows that connect the states. To activate a transition, a set of boolean conditions must be met.

\begin{figure}[h!]
    \centering
    \includegraphics[width=0.9\linewidth]{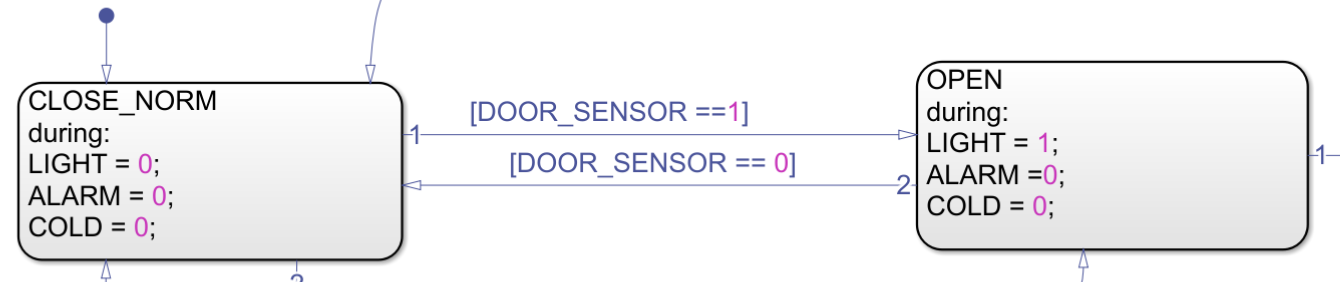}
    \caption{Extraction from a Stateflow diagram of a fridge temperature controller}
    \label{fig:stateflow}
\end{figure}

As an example of a Stateflow model, Figure~\ref{fig:stateflow} shows an excerpt of a Stateflow model that controls the temperature of a fridge. The example has four states, but only two are shown in the image: (1) CLOSE\_NORM indicates that the system door is closed and the temperature is cold enough not to activate the cool-down function (not in the image). If the door is opened (i.e., $DOOR\_SENSOR ==1 $), the system enters the state (2) OPEN. When this happens, the cool-down function is disabled, and a light is turned on through the variable LIGHT. In addition to these two states, the remaining two states, not shown in the figure, interact with these through transitions that represent additional operating conditions of the refrigerator.

\subsection{Mutation Analysis of Simulink Models}

Mutation analysis~\cite{budd1980mutation} aims at assessing a test suite adequacy by introducing small, systematic modifications, referred to as \textit{mutations}, into the system under test. In the context of Simulink-Stateflow models, these mutations are applied to various components of the model, including states, transitions and junctions, with the objective of simulating potential faults. A test suite is considered effective if it is capable of detecting these modifications; conversely, undetected mutations indicate areas of the model that may be insufficiently tested or prone to faults.

Mutations in Simulink-Stateflow can be systematically categorized based on the component they target. State-level mutations involve modifying state parameters or altering internal actions, such as replacing a \textit{during} action with an \textit{exit} action, thereby simulating common implementation faults. Transition-level mutations affect the connections between states, for example, by changing transition conditions, triggers, or target states, reflecting logical errors in state machine behavior. As illustrated in Figure~\ref{fig:stateflow}, one example of a transition-level mutation is changing the condition of a transition from $DOOR\_SENSOR == 1$ to $DOOR\_SENSOR \neq 1$, which alters the state machine’s behavior. Similarly, a state-level mutation could involve modifying the parameters of the \textit{CLOSE\_NORM} state, such as changing the value of $ALARM$ from 0 to 1. 

%\pablo{Update this part to show how mutation is composed}

%Mutation testing for Simulink models extends beyond traditional software testing applications. Mutations in Simulink are frequently utilized to mimic real-world faults, enabling tasks such as fault localization~\cite{liu2016simulink, arrieta2018multi}, test prioritization~\cite{matinnejad2018test, arrieta2019search}, and automated model repair~\cite{arrieta2024search}. These tasks benefit from high-quality mutations, which mimic real-world bugs in syntax and behavior. However, mutation testing in the context of Simulink models is complex due to the long execution times of the test cases and the model's complexity. In addition, redundant or non-executable mutants, common in mutant generation approaches, worsen these challenges by unnecessarily increasing computational costs. Thus, effective mutation testing in Simulink is paramount, it not only requires generating meaningful mutants but also generating executable mutants.

\section{Approach}\label{sec:Approach}
\subsection{Automated Mutant Generation Process}

\begin{figure*}[ht!]
    \centering
    \includegraphics[width=0.9\linewidth, trim = 0 0 0 0]{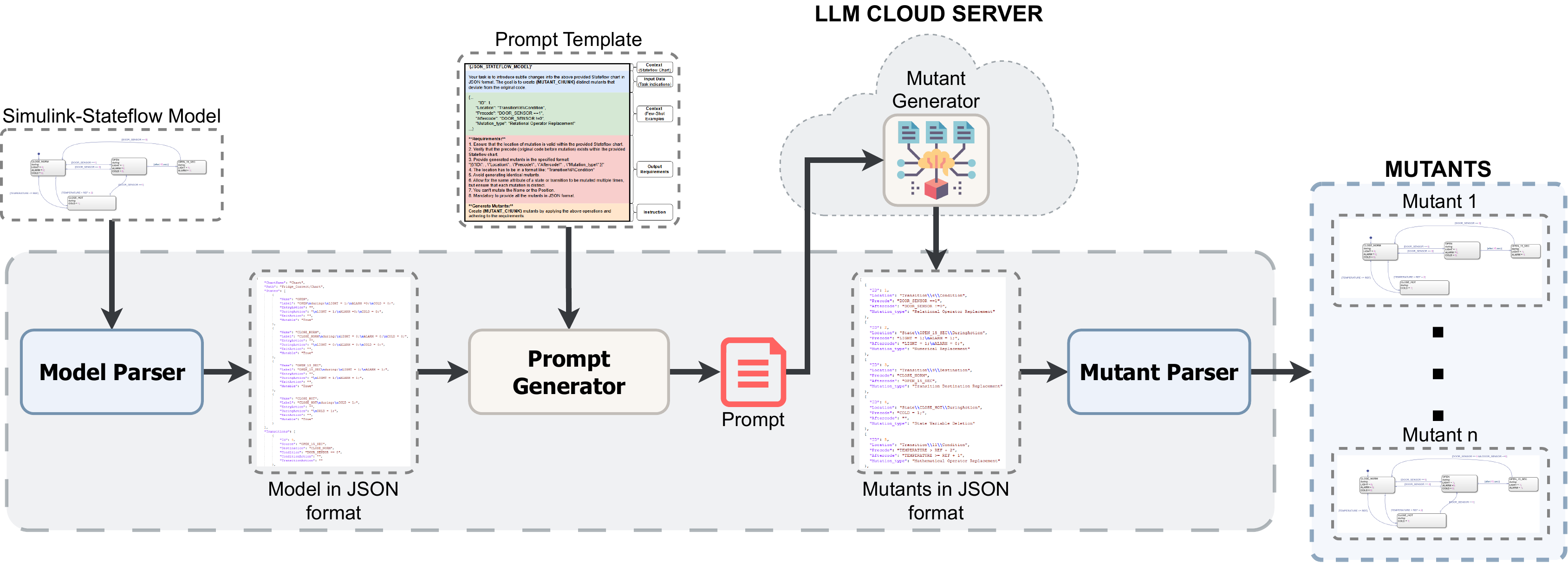}
    \caption{Architecture of the Automated Mutant Generation Process}
    \label{fig:architecture}
\end{figure*}

Figure \ref{fig:architecture} shows our automated mutant generation approach. The process begins with the Simulink model being parsed into a JSON format by the \textit{Model Parser}. This JSON representation includes the key elements of the Stateflow diagram within the Simulink model, such as States, Transitions, and Junctions. This structured format provides the foundational information to systematically apply mutations to the model.

Next, the \textit{Prompt Generator} takes the JSON representation and a predefined Prompt Template as inputs to create the prompt for a mutant generation. These prompts are constructed using a standardized template~\cite{wang2024comprehensive} as shown in Figure \ref{fig:prompt}, which provides an example of the prompt template used in our Mutant Generation Process. Once the prompts are generated, they are sent to an LLM cloud server. In our implementation, this server leverages either Ollama~\cite{OllamaAPI}, for open-source LLMs, or the OpenAI API~\cite{OpenAIAPI}, for GPT family closed-source LLMs. Then, the LLM processes the prompt and generates a response containing several mutants in JSON format. The number of generated mutants is specified in the Input Data section of the prompt. Finally, the \textit{Mutant Parser} converts the JSON mutants back into Simulink models, producing a Simulink model for each mutant proposed by the LLM. 

In our Automated Mutant Generation process, we define two distinct mutation strategies to control the scope of changes applied to the model. The Global mutation strategy, grants the LLM full access to the entire model. In this setting, the LLM may introduce modifications to any component without restriction. In contrast, the Local mutation strategy introduces a more focused and controlled form of mutation. Before the mutation is performed, a specific element of the model, such as a State, Transition, or Junction, is explicitly selected. The LLM is then instructed to apply mutations only within this predefined element.
% Figure \ref{fig:architecture} shows our automated mutant generation approach. At first, the Simulink model is taken and parsed into a JSON format by the \textbf{Model Parser}. The model JSON representation includes information about the Stateflow Chart inside the SImulink Model (i.e., States, Transitions and Junctions). This JSOn structure provides the information needed to make mutations of the model.

% Then, the \textit{Prompt Generator} which takes as input the Pormpt Template and the JSON representation of the model generates the prompts designed for the mutant generation. Prompts are buiilt with a set template to guarantee clear, consistent and relevant content. Figure \ref{fig:prompt} shows an  example of the prompt template for the Global mutation Strategy. After that, the prompt is sent to the LLM cloud Server, which in our case, is a server containing Ollama \pablo{Ref and explain} for open-source LLMs or the OpenAI API \pablo{ref and explain} for the closed-source LLMs. The LLM produces a response, which is analyzed and extracted the mutations in JSON format. Finally, the mutant parser takes the JSON of mutants and generates Simulink model for each mutant proposed by the LLM.

\subsection{Prompt Generation}

Prompts are crucial for guiding LLM responses, shaping the quality, relevance, and accuracy of the generated output. A well-designed prompt improves the LLM's performance on a specific task by providing clear direction and background. To guarantee the prompts are effective, we adhere to the best practices~\cite{guo2024exploring}, which suggest including four key elements: (1) Context, (2) Input Data, (3) Output Requirements, and (4) Instruction.

\begin{figure*}[ht!]
    \centering
    \includegraphics[width=0.60\linewidth, trim = 0 0 0 0]{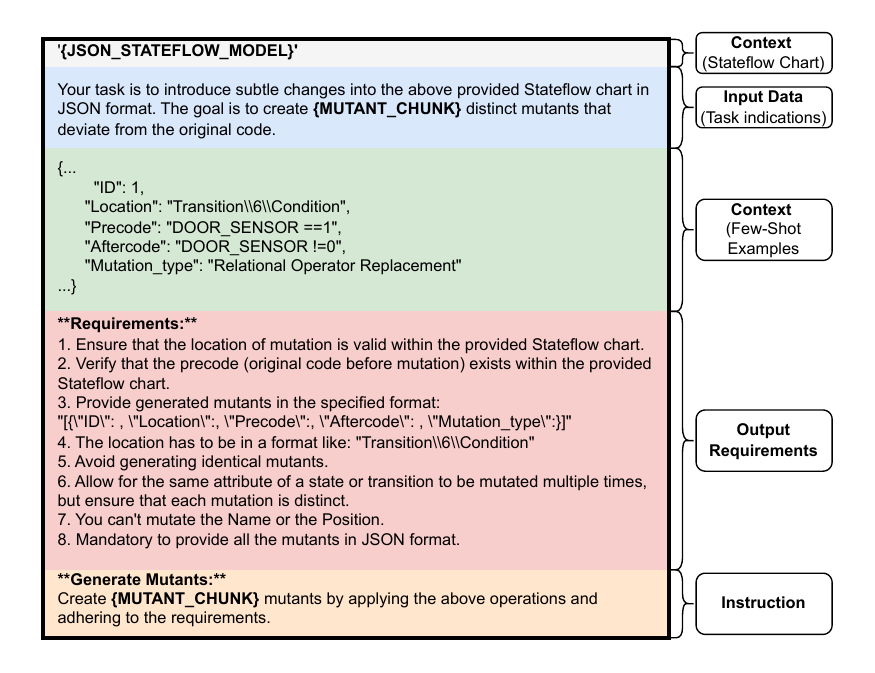}
    \caption{Prompt Template}
    \label{fig:prompt}
\end{figure*}

Figure \ref{fig:prompt} shows that the \textit{Context} section provides details about the object to be mutated to enhance the LLM's performance. This includes presenting the Stateflow model in JSON format and including examples for the few-shot learning approach. For few-shot examples, we used some real examples we selected from the baseline mutation algorithm (See Section~\ref{subsec:baseline} for further details). Next, the \textit{Input Data} section specifies both the model element to be mutated and the number of mutants to generate. Figure~\ref{fig:prompt} illustrates the prompt template used in the Global Mutation Strategy. In contrast, the Local Mutation Strategy only incorporates the specific State, Transition or Junction selected for mutation. For transitions, in addition to information about the transition itself, the IDs of the States or Junctions that may act as the source or destination are also included. For states, only the state itself is included, since states do not rely on other model elements. In the \textit{Output Requirements} section, we specify the JSON format for mutation outputs and provide several guidelines about the mutant generation, e.g., that attributes such as Name and Position must not be mutated. Finally, the prompt concludes with the \textit{Instruction}, where we explicitly direct the LLM to generate $N$ mutants based on the provided context, input data, and output rules.

% The \textit{Input Data} section specifies both the model element to be mutated and the number of mutants to generate. Figure~\ref{fig:prompt} illustrates the prompt template used in the Global Mutation Strategy. In contrast, the Local Mutation Strategy only incorporates the specific State or Transition selected for mutation. For transitions, in addition to information about the transition itself, the IDs of the States or Junctions that may act as the source or destination are also included. For states, only the state itself is included. Finally, in the \textit{Output} section, we specify the JSON format for mutation outputs and provide several guidelines about the mutant generation, for example, that attributes such as Name and Position must not be mutated. 

\section{Empirical Evaluation}\label{sec:Evaluation}

In our study, we evaluate the potential of LLMs in generating Simulink-Stateflow mutants by answering the following Research Questions (RQs):

\begin{itemize}
    \item \textit{\textbf{RQ1 -- Cost-Effectiveness:} How do LLMs perform in generating Simulink-stateflow Mutants in terms of efficiency and effectiveness?}
      
    \item \textit{\textbf{RQ2 -- Mutation Strategy Impact:} How does the mutation strategy affect LLMs' performance in generating mutants?}

    \item \textit{\textbf{RQ3 -- Prompt Strategy Impact:} To what extent does the prompt style affect the performance of LLMs generating mutants?}

    \item \textit{\textbf{RQ4 -- Mutation Diversity:} How does the configuration of the LLMs, specifically the temperature, affect the diversity and quality of generated mutants?}

    \item \textit{\textbf{RQ5 -- Generation Errors:} What are the primary causes of generation errors in non-compilable and non-generable Simulink-Stateflow mutants generated by LLMs?}
    
\end{itemize}

RQ1 aims at assessing whether LLMs have the potential to effectively and efficiently generate Simulink-Stateflow Mutants. To this end, we implemented a baseline algorithm and compared LLMs' performance against it. RQ2 aims at comparing the local mutation strategy with the global mutation strategy. RQ3 studies whether there is a performance difference between the zero-shot and few-shot prompting strategies. RQ4 aims to assess whether LLMs' temperature parameter affects the quality of the generated mutants. Lastly, in RQ5, we aim at analyzing the different root causes of compilation errors in the non-compilable generated mutants.

\subsection{Case studies} \label{sec:CaseStudies}

For our evaluation, we analyzed four case study systems, summarized in Table \ref{table:caseStudies}. These systems are modeled using Simulink-Stateflow and exhibit diverse characteristics in terms of complexity, size, and structure. Table \ref{table:caseStudies} shows key attributes such as the number of inputs, outputs, states, and transitions, which indicate the models' diversity.

%For our evaluation we used in total 4 case study systems. Table \ref{table:caseStudies} shows the different characteristics of the Stateflow models used in this evaluation. As can be seen, there are different size models \pablo{Add some interesting characteristics for comparison.} 

\begin{table}[ht!]
\caption{Characteristics of the Case Study Systems} 
\label{table:caseStudies}
\centering
\resizebox{0.47\textwidth}{!}{
\begin{tabular}{lcccc}
\toprule
 & \textbf{\# of Inputs} & \textbf{\# of Outputs} & \textbf{\# of States} & \textbf{\# of Transitions} \\ \cmidrule{1-5}
\textbf{Door} & 6 & 1 & 5 & 10 \\
\textbf{Fridge} & 3 & 3 & 4 & 8 \\
\textbf{Elevator} & 1 & 3 & 5 & 12 \\
\textbf{Pacemaker} & 12 & 15 & 11 & 17 \\ \bottomrule
\end{tabular}}
\end{table}

The first case study system, (\textit{Door}), controls an automated door. When it detects people, it automatically opens the door and later closes it. This is a relatively simple model as it only has five states and 10 transitions, enabling us to assess our approach's performance for simple systems. The second case study system, \textit{Fridge}, is a controller for a fridge with similar number of states and transitions as the \textit{Door} model. The third case study system is an elevator system that features fewer inputs (i.e., only one input) but a larger number of transitions (i.e., 12), reflecting a more dynamic behavior. We took the last case study system, i.e., \textit{Pacemaker} model, from the buggy models used by Ayesh et al.~\cite{ayesh2022two}. This is the most complex model, with a significantly larger number of inputs, outputs, states, and transitions. The diverse characteristics of these four stateflow models enable us to comprehensively evaluate our approach. Note that the four models were taken from prior work on repairing Simulink-Stateflow models~\cite{arrieta2024search}. %Note that the first three models were extracted from students' projects where they had to develop a Stateflow controller for a given situation. This practice of using students' projects is common in other areas, such as in automated program repair studies~\cite{gissurarson2022propr}.

To assess the quality of the generated mutants (see Section \ref{sec:evaluationMetrics} for further details), we asked an experienced engineer to develop a test suite of 10 to 16 test cases for every Simulink model, achieving at least 95\% decision coverage and 100\% execution coverage. Coverage~\cite{andrews2006using, cai2005effect} is a widely recognized metric for assessing test effectiveness, ensuring that the suite can reliably detect faulty behavior and increase the likelihood of identifying mutants that alter the system’s logic. Notice that in Simulink-Stateflow models, the number of test cases is usually limited because each test takes a long time to execute~\cite{valle2022towards}. For instance, one test case for the Pacemaker model includes signals lasting 300 seconds. We used mutation testing to compare the behavior of the generated mutants with the original model by executing the test suite. As explained in Section \ref{sec:background}, when conducting mutation testing, as an  oracle we compare the differences between the original model and the proposed mutant to determine if each mutant is detected or is still alive. This categorization enabled us to determine the quantity of test cases that effectively detected each mutant.

\subsection{Selected Large Language Models}
In our evaluation, we focus on the automated generation of Simulink-Stateflow mutants to comprehensively assess the performance of LLMs in this task. We incorporate closed-source models from the GPT family, including GPT-3.5-Turbo~\cite{brown2020language}, GPT-4o~\cite{openai2024gpt4}, and GPT-4o-mini~\cite{openai2024gpt4}. Moreover, following recent studies on evaluating the performance of LLMs in code-generation~\cite{liu2023your, jiang2024survey} we selected the most popular open-source models, including the 22-billion-parameter version of Mistral-Small~\cite{jiang2023mistral}, the 16-billion-parameter version of Deepseek-Coder-v2~\cite{zhu2024deepseek}, the 70-billion-parameter version of Llama~3~\cite{metallam60:online}, the 27-billion-parameter version of Gemma~\cite{team2024gemma}, and the 8-billion parameter version of Hermes~ \cite{teknium2024hermes3technicalreport}. More details about the models are shown in Table \ref{tab:modelChars}\footnote{OpenAI has not officially released the parameter size of their closed-source models.}. 

\begin{table}[h!]
\centering
\caption{Characteristics of the studied LLMs}
\label{tab:modelChars}
\resizebox{0.45\textwidth}{!}{
\begin{tabular}{lllrrr}
\toprule
\multicolumn{1}{c}{\textbf{\begin{tabular}[c]{@{}c@{}}Model \\ Type\end{tabular}}} & \multicolumn{1}{c}{\textbf{Model}} & \multicolumn{1}{c}{\textbf{\begin{tabular}[c]{@{}c@{}}Base \\ Model\end{tabular}}} & \multicolumn{1}{c}{\textbf{\begin{tabular}[c]{@{}c@{}}Training \\ Data Time\end{tabular}}} & \multicolumn{1}{c}{\textbf{\begin{tabular}[c]{@{}c@{}}Release \\ Time\end{tabular}}} & \multicolumn{1}{c}{\textbf{Size}} \\ \cmidrule{1-6}
\textbf{} & GPT-3.5-Turbo & GPT & 2021/09 & 2023/03 & - \\
\textbf{Closed} & GPT-4o & GPT & 2023/10 & 2024/05 & - \\
\textbf{} & GPT-4o-Mini & GPT & 2023/10 & 2024/07 & - \\ \cmidrule{1-6}
 & Mistral-Small & Mistral & 2023/10 & 2024/09 & 22B \\
 & Deepseek-Coder-v2 & DeepSeek & 2023/09 & 2024/07 & 16B \\
\textbf{Open} & Llama 3 & Llama & - & 2024/04 & 70B \\
\textbf{} & Gemma 2 & Gemini & - & 2024/06 & 17B \\
 & Hermes 3 & Llama 3.1 & 2024/08 & 2024/08 & 8B \\ \bottomrule
\end{tabular}}
\end{table}

\subsection{Baseline Algorithm} \label{subsec:baseline}

To assess the effectiveness of LLMs in RQ1, we implemented a baseline mutation algorithm (See Algorithm~\ref{alg:MutationAlgorithm}). This algorithm takes the original Stateflow model as input and generates a mutated version by applying mutation operators specifically designed for Simulink-Stateflow models, as proposed by Arrieta et al.~\cite{arrieta2024search} in their search-based repair study. In total, the operator set includes 8 operators for states and 11 for transitions. To ensure a fair comparison with LLMs, which can introduce more than one mutation within a single mutant, we generated two sets of mutants. First, we generated 100 first-order mutants by applying only one mutation (i.e., the probability of mutating more than once is zero, Algorithm ~\ref{alg:MutationAlgorithm} line 12). Second, we used the baseline algorithm to generate additional 100 mutants which had the possibility of being higher-order mutants (i.e., multiple mutations within a single mutant). The mutation process for the second set of mutants follows a probabilistic stopping condition, since after each mutation, the likelihood of performing an additional one decreases exponentially as $0.5^{numOfMutations}$. In each iteration, the algorithm randomly decides whether to mutate a state or a transition (Algorithm~\ref{alg:MutationAlgorithm} Line 5). Then the algorithm applies the corresponding mutation operators either for the transitions or the states (Algorithm~\ref{alg:MutationAlgorithm} Lines 6-9). The target element (i.e., state or transition) is randomly selected, followed by the random selection of a valid mutation operator. If the chosen operator is not applicable to the selected element (i.e., state or transition), the algorithm retries with another operator until a valid operator is found. This is because some operators are not applicable to all states or transitions. %For example, the Conditional Replacement Operator, which replaces the conditional operator of a statement, cannot be applied to a transition that does not contain a conditional statement.

\begin{algorithm}[ht]
\caption{Baseline Stateflow mutant generation algorithm}\label{alg:MutationAlgorithm}
   % \begin{algorithmic}

% Start a box that is 80% (0.8) of the text width

    \KwIn{model  //\textit{Path to original model} }
    \KwOut{mutModel  //\textit{Path to mutated model} \\}

    $mutModel$ $\leftarrow$ copyModel($model$)
    
    $numOfMutations$ $\leftarrow$ 0
    
    [$states, transitions$] $\leftarrow$ getAttributes($mutModel$)
    
    \Do{$p < 0.5^{numOfMutations}$}{
      $selection$ $\leftarrow$ randSelTransState()

      \If{$selection == Transition$}{
         $mutModel$ $\leftarrow$ mutateTransition($mutModel$)
       }
       \Else{
         $mutModel$ $\leftarrow$ mutateState($mutModel$)
       }

      $numOfMutations$ $\leftarrow$ $numOfMutations +1$ 
      
      $p$ $\leftarrow$ rand()
    }

    \Return $mutModel$
   % \end{algorithmic}

\end{algorithm}
\vspace{-0.4cm}
 
\subsection{Experimental Settings} 
\subsubsection{Settings for RQ1} To investigate the performance of LLMs to generate mutants in terms of efficiency and effectiveness metrics, we compared 25 generated mutants by each LLM across different mutation, prompt, and model configurations. In total, we evaluated 25 mutants $\times$ 4 Simulink models $\times$ 2 mutation strategies $\times$ 4 prompt strategies $\times$ 6 LLM temperature values $\times$ 8 LLMs = 38,400 mutants in total, 4800 mutants per model. In addition, we adopted a baseline algorithm (see Section \ref{subsec:baseline}) to generate 200 mutants per Simulink model, i.e., a total of 800 mutants.

\subsubsection{Settings for RQ2 and RQ3} We explored how different mutation and prompt strategies affect the performance of mutant generation. To do so, we generated two sets of prompt templates (i.e., one for the mutation strategy and one for the prompt strategy). More specifically, we generated 4 prompts by combining both sets of prompt templates by adding or removing different sources of information in the context shown as follows:
\begin{enumerate}
    \item \textbf{Prompt 1 (P1): The default Prompt.} P1 is the default prompt as shown in Figure \ref{fig:prompt}, but without the Few-Shot examples.
    \item \textbf{Prompt 2 (P2): P1 with selected state or transition.} P2 is based on P1 but tailored for the local mutation strategy, specifying the selected state or transition to be mutated.
    \item \textbf{Prompt 3 (P3): P1 with Few-Shot examples.} P3 builds on P1 by incorporating several mutation examples (i.e., Few-Shot examples) as shown in Figure \ref{fig:prompt}. 
    \item \textbf{Prompt 4 (P4): P2 with Few-Shot examples.} P4 expands on P2 by including several mutation examples. 
\end{enumerate}

In addition to these four templates, for RQ3, we also evaluated the impact of different few-shot example numbers (i.e., 3 examples, 6 examples and 9 examples) as well as the comparison against the zero-shot prompt template. The choice of these few-shot example numbers was guided by prior work~\cite{deng2024large, xie2021explanation}, which has shown that varying the number of examples can meaningfully affect model behavior.

\subsubsection{Settings for RQ4} To understand how the configuration of the LLM creativity setting affects the diversity and the quality of the generated mutants, we evaluated all models with 6 different temperature values (i.e., 0.2, 0.4, 0.6, 0.7, 0.8, 1.0). These values span the full range of possible temperatures (i.e., 0 to 1), with 0.6–0.8 representing commonly used default values, allowing us to assess performance under both conservative and more creative settings. For each temperature, we evaluated 6400 mutants by assessing the effectiveness metrics of each set of mutants.

\subsubsection{Settings for RQ5} Providing non-generable and non-compilable mutants increases the cost of the mutant generation process. Thus, this RQ studies the root causes of generating such mutants. Specifically, we distinguish between two types of problematic mutants: 1) non-generable mutants, for which the mutant parser of our approach cannot generate any Simulink-Stateflow model; and 2) non-compilable mutants, which are generated but cannot be compiled by Simulink. We selected mutants in both categories and analyzed the underlying reasons for these failures, either at the generation stage or during compilation.% we selected the non-compilable and non-generable mutants proposed by each LLM. We analyzed the reasons they were not being compiled by the Simulink compiler or generated by the mutant parser. 

\subsection{Evaluation Metrics} \label{sec:evaluationMetrics}
%\pablo{Add Table to match RQs and evaluation metrics}
To assess the effectiveness and efficiency of different LLMs when generating Stateflow mutants, we used a set of metrics proposed by Wang et al. \cite{wang2024exploratory}. To do so, we differentiate two categories of evaluation metrics capturing different aspects: 

\subsubsection{Effectiveness Metrics}
The effectiveness metrics indicate the ability of the LLMs to generate mutants:

\noindent\textbf{Mutant count:} This metric measures the total number of generated mutants. It is relevant for evaluating LLM effectiveness because the number of produced mutants does not always match the requested amount. For instance, when asked to generate 100 mutants, an LLM may output fewer (or occasionally more) due to limitations in following precise quantitative instructions. Thus, deviations in mutant count provide insight into the reliability of the model.

\noindent\textbf{Generation Rate:} LLMs may suffer from hallucination and propose non-generable mutants (e.g., a change in the source of one transition to a non-existing state, making it impossible for the parser module in our approach to generate the Simulink-Stateflow model). Therefore, with this metric, we measure the proportion of generable mutants out of the total number of mutants proposed by an LLM.

\noindent\textbf{Compilation Rate:} Even when the generation of a proposed mutant is correct (i.e., the mutant parser in our approach succeeds in generatingto generate the Simulink-Stateflow model), not all generated mutants are guaranteed to compile. This metric measures the proportion of mutants that successfully compile after generation.

% \noindent\textbf{Useless Mutant Rate:} LLMs may produce duplicated mutants, or the mutants identical to the original Stateflow model. Sometimes, LLMs struggle with issues of misunderstanding and repetition, leading to an ineffective mutant generation. With this metric, we measure the proportion of the mutants that are duplicated or identical to the original model. 

\noindent\textbf{Duplication Rate:} LLMs may generate mutants that are identical or highly similar. To quantify this, we measured the proportion of proposed mutants that exhibit such similarity. Our evaluation involved a two-step verification process. First, we compared the test suite verdicts for each mutant to identify those detected by the same test cases. Next, for mutants detected by the same tests, we compared their outputs to assess whether they exhibited equivalent behavior.

\noindent\textbf{Equivalent Mutant Rate:}
Similar to the duplication rate, LLMs may generate mutants semantically identical to the original one. To quantify this, we used the Equivalent Mutant definition proposed by Estero et al.~\cite{estero2015quality} which measures the proportion of mutants equivalent to the original mutant. This evaluation was conducted in two steps: first, we compared the outputs of each mutant with those of the original model; then, for mutants exhibiting similar behavior (i.e., Potentially Equivalent Mutants), we manually assessed their semantic equivalence with respect to the original model. Given the large number of Potentially Equivalent Mutants, a complete manual validation was infeasible. Therefore, we manually inspected a statistically representative sample of 246 mutants in total. The representative sample size was determined using Cochran’s Sample Size formula, where we set the confidence level to be 95\%, the margin of error to be 5\%, and the estimated population proportion to 0.2, which corresponds to the fraction of mutants deemed potentially equivalent.

%\noindent\textbf{Killability rate:} \pablo{maybe we can remove this since we are using the quality of the mutants}

\noindent\textbf{Quality of the generated mutants:} The metric proposed by Estero et al.~\cite{estero2015quality} evaluates the quality of generated mutants in relation to a given test suite. It correlates the number of test cases in a test suite that successfully detect a mutant with the number of other mutants each test case detects. A higher value for this metric indicates a more challenging mutant, as it is detected by fewer test cases that do not also detect other mutants. %To comprehensively evaluate a given test strategy, it is important to have mutants with different degrees of quality (e.g., some easy to detect and some difficult to detect).

% \noindent\textbf{Hallucination Rate ($H_r$):} Huge efforts have been made to measure the hallucination in LLMs in different contexts \cite{ji2023survey, rawte2023troubling, ji2023towards}. However, to the best of our knowledge, no metric exists for measuring hallucination in our context. Therefore, we defined the Hallucination Rate, which quantifies the extent to which each LLM hallucinates within our specific context. This metric is derived from a combination of the Generation rate ($G_r$) and Compilation rate ($C_r$) and is calculated as follows:

% \begin{equation}
%     H_r = 1 - \frac{G_r + C_r}{2}
% \end{equation}

% \noindent Therefore, the lower this metric, the lower the hallucinations faced by the LLM.

% \noindent\textbf{Usable Mutant rate:} This metric measures the ratio between the generated mutant and the usable mutants. A mutant is considered usable if it is compilable, not equivalent to the original model, and not a duplicate of another mutant. A higher value for this metric indicates a greater likelihood that the model will produce usable mutants for the end-user.

\subsubsection{Efficiency Metrics}
The Efficiency Metrics indicate the cost of generating each mutant in terms of money and time:

\noindent\textbf{Generation time:} This metric measures the time in seconds it takes for an LLM to generate a mutant.

\noindent\textbf{Generation cost:} This metric measures the cost in euros to generate one mutant.

% \begin{table}[ht]
% \centering
% \caption{Overview of the Evaluation Metrics used across Research Questions}
% \label{tab:my-table}
% \begin{tabular}{llllll}
% \toprule
% \multicolumn{1}{c}{\textbf{}} & \multicolumn{1}{c}{\textbf{$RQ_1$}} & \multicolumn{1}{c}{\textbf{$RQ_2$}} & \multicolumn{1}{c}{\textbf{$RQ_3$}} & \multicolumn{1}{c}{\textbf{$RQ_4$}} & \multicolumn{1}{c}{\textbf{$RQ_5$}} \\ \cmidrule{1-6}
% \textbf{Mutant Count} & + & + & + & + & - \\
% \textbf{Generation Time} & + & + & - & - & - \\
% \textbf{Generation Cost} & + & - & - & - & - \\
% \textbf{Generation Rate} & + & + & + & + & - \\
% \textbf{Compilation Rate} & + & + & + & + & - \\
% \textbf{Duplication Rate} & + & + & + & + & - \\
% \textbf{Equivalent Mutant Rate} & + & + & + & + & - \\
% \textbf{Quality of the mutants} & + & + & + & + & - \\
% %\textbf{Hallucination Rate} & - & - & - & + & + \\ 
% \bottomrule
% \end{tabular}
% \end{table}

\subsection{Execution Environment}
We used MATLAB 2022b for the Simulink models and the generated mutants. For the prompt generation and model inference, we used Python 3.10.11. To execute the code for our approach, we used a Windows 11 computer with 32 GB RAM memory and a $12^{th}$ Intel Core i5-1235U processor with 10 cores and 12 threads. To execute the LLMs apart from the OpenAI API~\cite{OpenAIAPI} for the closed-source LLMs, we used a Linux server with 128 GB of RAM, an AMD EPYC 7773X 64-Core Processor, and an NVIDIA RTX A6000 with 48 GB of VRAM.

\section{Analysis of the results and Discussion}\label{sec:Discussion}

In this section we present a detailed analysis of the experimental results corresponding to our research questions.

\subsection{$RQ_1$ - Cost-Effectiveness}

% \usepackage{multirow}
\begin{table*}[ht!]
\centering
\caption{$RQ_1$ - Cost-Effectiveness comparison of the evaluated LLMs. Best results are marked in bold}
\label{tab:RQ1}
\resizebox{0.80\textwidth}{!}{
\begin{tabular}{clrrrrrrrrr}
\toprule 
 &  & \multicolumn{1}{l}{GPT 3.5} & \multicolumn{1}{l}{GPT 4o} & \multicolumn{1}{l}{GPT 4o-m} & \multicolumn{1}{l}{DSC-V2} & \multicolumn{1}{l}{Hermes 3} & \multicolumn{1}{l}{Llama 3} & \multicolumn{1}{l}{Gemma 2} & \multicolumn{1}{l}{Mistral-S} & \multicolumn{1}{l}{Baseline} \\ \cmidrule{1-11}
\multirow{3}{*}{Cost (€)} & Total & 0.8200 & 3.9900 & 0.2700 & \multicolumn{1}{r}{\textbf{0.0000}} & \multicolumn{1}{r}{\textbf{0.0000}} & \multicolumn{1}{r}{\textbf{0.0000}} & \multicolumn{1}{r}{\textbf{0.0000}} & \multicolumn{1}{r}{\textbf{0.0000}} & \multicolumn{1}{r}{\textbf{0.0000}}\\
 & Unitary & 0.0002 & 0.0008 & \textless{}0.0001 & \multicolumn{1}{r}{\textbf{0.0000}} & \multicolumn{1}{r}{\textbf{0.0000}} & \multicolumn{1}{r}{\textbf{0.0000}} & \multicolumn{1}{r}{\textbf{0.0000}} & \multicolumn{1}{r}{\textbf{0.0000}} & \multicolumn{1}{r}{\textbf{0.0000}}  \\
 & Prompt & 0.0043 & 0.0208 & 0.0014 & \multicolumn{1}{r}{\textbf{0.0000}} & \multicolumn{1}{r}{\textbf{0.0000}} & \multicolumn{1}{r}{\textbf{0.0000}} & \multicolumn{1}{r}{\textbf{0.0000}} & \multicolumn{1}{r}{\textbf{0.0000}} & \multicolumn{1}{r}{\textbf{0.0000}} \\ \cmidrule{1-11}
 
\multirow{4}{*}{\begin{tabular}[c]{@{}c@{}}Generation \\ time (s)\end{tabular}} & Door & \textbf{0.510} & 0.980 & 1.180 & 1.060 & 0.600 & 4.640 & 2.510 & 1.960 & 6.660 \\
& Elevator & \textbf{0.660} & 1.270 & 0.940 & 1.170 & 0.800 & 5.350 & 2.900 & 2.110 & 4.520 \\
& Fridge & \textbf{0.540} & 1.160 & 0.970 & 1.170 & 0.630 & 4.980 & 2.580 & 1.990 & 3.020 \\
& Pacemaker & 1.550 & 1.250 & 1.450 & 1.680 & \textbf{0.970} & 5.100 & 4.580 & 4.110 & 5.430 \\ \cmidrule{1-11}
\multirow{4}{*}{\begin{tabular}[c]{@{}c@{}}Mutant \\ Count\end{tabular}} & Door & 0.979 & \textbf{1.000} & \textbf{1.000} & 0.958 & 0.983 & 0.920 & 0.995 & 0.941 & \textbf{1.000} \\
& Elevator & 0.995 & \textbf{1.000} & \textbf{1.000} & 0.941 & 0.933 & 0.925 & 0.937 & 0.941 & \textbf{1.000} \\
& Fridge & \textbf{1.000} & \textbf{1.000} & \textbf{1.000} & 0.933 & 0.975 & 0.920 & 0.979 & 0.966 & \textbf{1.000} \\
& Pacemaker & 0.870 & \textbf{1.000} & \textbf{1.000} & 0.937 & 0.995 & 0.962 & 0.895 & 0.941 & \textbf{1.000} \\ \cmidrule{1-11}
\multirow{4}{*}{\begin{tabular}[c]{@{}c@{}}Generability \\ Rate\end{tabular}} & Door & 0.911 & 0.944 & 0.964 & 0.850 & 0.827 & 0.929 & 0.931 & 0.878 & \textbf{1.000} \\
& Elevator & 0.971 & 0.994 & 0.923 & 0.914 & 0.918 & 0.957 & 0.965 & 0.931 & \textbf{1.000} \\
& Fridge & 0.972 & 0.984 & 0.966 & 0.919 & 0.926 & 0.933 & 0.942 & 0.955 & \textbf{1.000} \\
& Pacemaker & 0.924 & \textbf{1.000} & 0.970 & 0.836 & 0.916 & 0.951 & 0.838 & 0.880 & \textbf{1.000} \\ \cmidrule{1-11}
\multirow{4}{*}{\begin{tabular}[c]{@{}c@{}}Compilability \\ Rate\end{tabular}} & Door & 0.745 & 0.730 & 0.763 & 0.557 & 0.550 & 0.640 & 0.595 & 0.611 & \textbf{0.920} \\
& Elevator & 0.861 & 0.895 & 0.823 & 0.653 & 0.694 & 0.829 & 0.786 & 0.782 & \textbf{0.985} \\
& Fridge & 0.839 & 0.827 & 0.804 & 0.700 & 0.788 & 0.692 & 0.682 & 0.768 & \textbf{0.980} \\
& Pacemaker & 0.822 & 0.840 & 0.786 & 0.551 & 0.633 & 0.603 & 0.598 & 0.698 & \textbf{0.930} \\ \cmidrule{1-11}
\multirow{4}{*}{\begin{tabular}[c]{@{}c@{}}Duplication \\ Rate\end{tabular}} & Door & 0.505 & 0.406 & 0.413 & 0.325 & 0.329 & 0.351 & \textbf{0.308} & 0.327 & 0.670 \\
& Elevator & 0.567 & \textbf{0.130} & 0.172 & 0.277 & 0.303 & 0.189 & 0.149 & 0.219 & 0.390 \\
& Fridge & 0.278 & \textbf{0.165} & 0.188 & 0.355 & 0.338 & 0.176 & 0.207 & 0.278 & 0.500 \\
& Pacemaker & 0.485 & 0.070 & \textbf{0.041} & 0.226 & 0.393 & 0.113 & 0.131 & 0.141 & 0.340 \\ \cmidrule{1-11}
\multirow{4}{*}{\begin{tabular}[c]{@{}c@{}} Equivalent \\ Rate\end{tabular}} & Door & 0.194 & 0.028 & 0.024 & 0.152 & 0.181 & 0.043 & \textbf{0.013} & 0.050 & 0.085 \\
& Elevator & 0.477 & \textbf{0.017} & 0.041 & 0.250 & 0.245 & 0.079 & 0.067 & 0.104 & 0.035 \\
& Fridge & 0.154 & \textbf{0.025} & 0.071 & 0.289 & 0.244 & 0.069 & 0.083 & 0.152 & 0.040 \\
& Pacemaker & 0.457 & 0.027 & \textbf{0.013} & 0.228 & 0.269 & 0.057 & 0.046 & 0.059 & 0.065 \\ \cmidrule{1-11}
\multirow{4}{*}{\begin{tabular}[c]{@{}c@{}} Mutant \\ Quality\end{tabular}} & Door & 0.560 & 0.755 & \textbf{0.763} & 0.503 & 0.486 & 0.694 & 0.701 & 0.655 & 0.620 \\
& Elevator & 0.185 & 0.410 & 0.399 & 0.336 & 0.346 & 0.482 & 0.441 & \textbf{0.496} & 0.283 \\
& Fridge & 0.645 & \textbf{0.787} & 0.736 & 0.510 & 0.589 & 0.733 & 0.722 & 0.654 & 0.710 \\
& Pacemaker & 0.411 & 0.869 & \textbf{0.882} & 0.507 & 0.551 & 0.802 & 0.719 & 0.705 & 0.778 \\
 \bottomrule
\end{tabular}}
\end{table*}

Table~\ref{tab:RQ1} shows the overall results for RQ1. In terms of monetary cost, only the GPT family models required payment (as the rest of the models could be deployed in our GPU cluster). Among them, GPT 4o was the most expensive while GPT 4o-m was the cheapest. All other models were free to use but required dedicated hardware to run locally, which in practice can make them even more expensive than the GPT API. The baseline was the fully free one, since it required neither paid API usage nor specialized infrastructure. 

Regarding efficiency, all LLMs were significantly faster than the baseline, confirming a clear efficiency advantage of LLM-based generation over the baseline. The GPT family was the fastest, achieving 13x speedups, with GPT 3.5 being the fastest model overall. Some open-source models, such as Llama 3, Gemma 2, and Mistral-S, showed slower generation times among LLMs due to their larger model sizes.

With respect to the effectiveness, which involves not only how many mutants each LLM produced, but also the quality of the produced mutants, LLMs demonstrated to be effective at generating mutants. From a mutant generation perspective, the baseline performed the best, achieving a perfect generability (i.e., 1.0 in all the cases) and the highest compilability rates (i.e., up to 0.985). This indicates that the baseline produces highly valid mutants with very few syntactic errors. However, the baseline often generates many trivial or redundant mutants that add little testing value, as shown by the low mutant quality values. In contrast, LLMs were slightly less consistent in generability and compilability (i.e., typically in the 0.90-0.97 range), but produced mutants with higher behavioral diversity. LLMs generated fewer duplicated mutants compared to the baseline, reducing the duplication rate from 0.67 to 0.31 in the Door case study and from 0.34 to 0.04 in the Pacemaker case study. Moreover, LLMs produced significantly fewer equivalent mutants, for instance, Gemma 2 achieved almost 6 times fewer equivalent mutants than the baseline in the Door case study. Despite slightly lower generability and compilability rates, most LLMs achieved higher overall mutant quality than the baseline. GPT 4o and GPT 4o-m consistently ranked among the best performers, and even open-source models were competitive.

\begin{custombox}{Answer to RQ1}
LLMs demonstrated to be both effective and efficient at generating mutants as they generate mutants up to 13x faster than the baseline. Although not all the mutants generated by LLMs are syntactically correct, LLMs generated higher-impact mutants, with greater mutant quality, fewer duplicates, and equivalent mutants compared to the baseline.
\end{custombox}

\subsection{$RQ_2$ - Mutation Strategy Impact} \label{Results:RQ2}

\begin{table*}[ht!]
\centering
\caption{$RQ_2$ - Comparison of the Global and Local mutation strategy impact in the mutant generation performance}
\label{tab:RQ2}
\resizebox{0.975\textwidth}{!}{
\begin{tabular}{lrr|rr|rr|rr|rr|rr|rr|rr}
\toprule
 & \multicolumn{2}{c}{\textbf{GPT 3.5}} & \multicolumn{2}{c}{\textbf{GPT 4o}} & \multicolumn{2}{c}{\textbf{GPT 4o-m}} & \multicolumn{2}{c}{\textbf{DSC-V2}} & \multicolumn{2}{c}{\textbf{Hermes 3}} & \multicolumn{2}{c}{\textbf{Llama 3}} & \multicolumn{2}{c}{\textbf{Gemma 2}} & \multicolumn{2}{c}{\textbf{Mistral-S}} \\ \cmidrule{2-17}
 & \textbf{Global} & \textbf{Local} & \textbf{Global} & \textbf{Local} & \textbf{Global} & \textbf{Local} & \textbf{Global} & \textbf{Local} & \textbf{Global} & \textbf{Local} & \textbf{Global} & \textbf{Local} & \textbf{Global} & \textbf{Local} & \textbf{Global} & \textbf{Local} \\ \cmidrule{1-17}
\textbf{Mutant Count} & 0.927 & \textbf{0.995} & \textbf{1.000} & \textbf{1.000} & \textbf{1.000} & \textbf{1.000} & \textbf{0.968} & 0.916 & \textbf{0.979} & 0.964 & \textbf{1.000} & 0.864 & 0.906 & \textbf{0.997} & \textbf{0.979} & 0.916 \\
\textbf{Generation Time (s)} & 0.794 & \textbf{0.787} & \textbf{1.040} & 1.291 & \textbf{0.963} & 1.310 & \textbf{0.867} & 1.689 & 0.853 & \textbf{0.647} & \textbf{3.864} & 6.359 & \textbf{2.834} & 3.360 & 2.667 & \textbf{2.399} \\
\textbf{Generability Rate} & \textbf{0.978} & 0.915 & \textbf{0.990} & 0.972 & \textbf{0.968} & 0.943 & \textbf{0.955} & 0.800 & \textbf{0.943} & 0.849 & \textbf{0.978} & 0.902 & 0.914 & \textbf{0.926} & \textbf{0.915} & 0.907 \\
\textbf{Compilability Rate} & \textbf{0.929} & 0.712 & \textbf{0.905} & 0.741 & \textbf{0.841} & 0.747 & \textbf{0.691} & 0.535 & \textbf{0.780} & 0.549 & \textbf{0.766} & 0.603 & \textbf{0.702} & 0.631 & \textbf{0.766} & 0.660 \\
\textbf{Duplication Rate} & 0.682 & \textbf{0.249} & \textbf{0.173} & 0.212 & \textbf{0.177} & 0.230 & 0.356 & \textbf{0.233} & 0.414 & \textbf{0.268} & \textbf{0.198} & 0.215 & \textbf{0.175} & 0.225 & 0.265 & \textbf{0.216} \\
\textbf{Equivalent Rate} & 0.605 & \textbf{0.048} & 0.029 & \textbf{0.019} & 0.043 & \textbf{0.031} & 0.283 & \textbf{0.172} & 0.267 & \textbf{0.202} & \textbf{0.060} & 0.064 & \textbf{0.051} & 0.052 & 0.116 & \textbf{0.065} \\
\textbf{Mutant Quality} & 0.269 & \textbf{0.632} & \textbf{0.714} & 0.696 & \textbf{0.705} & 0.684 & 0.455 & \textbf{0.473} & \textbf{0.518} & 0.468 & \textbf{0.704} & 0.652 & 0.602 & \textbf{0.689} & \textbf{0.639} & 0.616 \\
\bottomrule
\end{tabular}}
\end{table*}

To investigate the impact of mutation strategy on LLM-based mutant generation, we compared Global and Local mutation strategies across all models. As reported in Table~\ref{tab:RQ2}, the Global strategy was more time-efficient in five out of eight LLMs, likely due to prompt complexity. Local mutation required more computation to prepare the prompt (i.e., extracting all necessary features from the Simulink-Stateflow model), thus increasing the time needed to generate each mutant. %Additionally, while Global mutation allows changes across the entire model, generating valid and coherent mutants within a constrained Local context can be more challenging.

Regarding effectiveness, the total number of generated mutants did not consistently favor one strategy, varying across models. However, a clear trend emerges in generability and compilability. Global mutation consistently outperformed Local mutation in almost all cases, producing up to 20\% more syntactically correct mutants. This suggests that enabling mutations across the entire model enables LLMs to better maintain structural and variable dependencies, resulting in more valid and executable mutants. In contrast to this trend, duplication and equivalent mutant rates favored the Local strategy. For example, GPT-3.5 reduced its duplication rate from 0.682 (Global) to 0.249 (Local) and its equivalent mutant rate from 0.605 (Global) to 0.048 (Local). This high equivalent mutant rate for GPT-3.5 was caused due to the LLM not producing any mutation in certain mutants. Similar improvements in favour of the local mutation strategy were observed in DSC-V2, Hermes 3, and Mistral-S. Even where Global slightly outperformed Local, differences were minimal, at most 0.05, whereas Local’s improvements were substantial, at least 0.05 in most cases. This demonstrates that guiding the model to a specific location effectively reduces duplicated and equivalent mutants, a common issue in mutant generation approaches.

With respect to the mutant quality, there is no clear overall winner. GPT-3.5 notably benefited from Local mutation (0.632 vs. 0.269), while for most other models, differences were minimal (i.e., lower than 0.08). Overall, Global and Local strategies showed complementary strengths. Global excels at generating syntactically correct and compilable mutants, whereas Local produces more diverse and higher-quality mutants by reducing redundancy. These findings indicate that the choice of mutation strategy should be guided by the specific LLM. For GPT-4o, GPT-4o-m, Llama 3, Gemma 2, and Mistral-S, Global mutation is preferable, as the improvements offered by Local mutation in reducing duplication or equivalence were relatively small, while the Global strategy consistently achieved higher generability and compilability. In contrast, for GPT-3.5, DSC-V2, and Hermes 3, Local mutation is recommended, as it substantially reduces duplication and equivalent mutants, improving the overall quality of the generated mutants.

\begin{custombox}{Answer to RQ2}
The mutation strategy significantly affects LLM performance in generating mutants. Overall, Global mutation improves generability and compilability, while Local mutation reduces  duplicate and equivalent mutants. The optimal strategy depends on the LLM, as performance varies across different models.
\end{custombox}

\subsection{$RQ_3$ - Prompt Strategy Impact} \label{Results:RQ3}

Following the same trend observed in the previous research question, the results presented in Table~\ref{tab:RQ3} demonstrate that the prompt strategy substantially influences the performance of LLMs in generating mutants. Few-Shot prompts consistently outperformed zero-shot prompts across all evaluated metrics. Specifically, global Few-Shot prompts (P3) achieved the highest generability and compilability, enabling the generation of a larger number of valid and executable mutants. In contrast, local Few-Shot prompts (P4) excelled at producing diverse and high-quality mutants, significantly reducing duplication and equivalent mutants while improving overall mutant quality. This distinction between local and global approaches mirrors the patterns observed in the comparison of mutation strategies in the previous research question.

\begin{table}[ht!]
\centering
\caption{$RQ_3$ - Comparison of the Prompt Strategy Impact in the mutant generation performance}
\label{tab:RQ3}
\resizebox{0.48\textwidth}{!}{
\begin{tabular}{lr|r|rrr|rrr}
\toprule
 & \multicolumn{1}{c|}{\textbf{P1}} & \multicolumn{1}{c|}{\textbf{P2}} & \multicolumn{3}{c|}{\textbf{P3}} & \multicolumn{3}{c}{\textbf{P4}} \\ \cmidrule{2-9}
 & \multicolumn{1}{c|}{\textbf{Z-S}} & \multicolumn{1}{c|}{\textbf{Z-S}} & \multicolumn{1}{c}{\textbf{F-S 3}} & \multicolumn{1}{c}{\textbf{F-S 6}} & \multicolumn{1}{c|}{\textbf{F-S 9}} & \multicolumn{1}{c}{\textbf{F-S 3}} & \multicolumn{1}{c}{\textbf{F-S 6}} & \multicolumn{1}{c}{\textbf{F-S 9}} \\ \cmidrule{1-9}
\textbf{Mutant Count} & 0.963 & 0.946 & \textbf{0.979} & 0.973 & 0.963 & 0.960 & 0.954 & 0.966 \\
\textbf{Generability rate} & 0.932 & 0.818 & 0.961 & 0.959 & \textbf{0.971} & 0.925 & 0.946 & 0.922 \\
\textbf{Compilability rate} & 0.804 & 0.572 & \textbf{0.808} & 0.785 & 0.796 & 0.672 & 0.673 & 0.679 \\
\textbf{Duplication Rate} & 0.450 & 0.218 & 0.252 & 0.254 & 0.256 & \textbf{0.195} & 0.264 & 0.248 \\
\textbf{Equivalent Rate} & 0.376 & 0.121 & 0.109 & 0.114 & 0.120 & \textbf{0.051} & 0.069 & 0.082 \\
\textbf{Mutant Quality} & 0.415 & 0.539 & 0.627 & 0.639 & 0.623 & 0.659 & \textbf{0.673} & 0.584 \\
\bottomrule
\end{tabular}}
\end{table}

Zero-Shot local prompts (P2), which specify the mutation target without examples, consistently underperformed both in generability and compilability compared to the corresponding Few-Shot prompts (P4). This highlights the critical role of guidance when focusing on specific model fragments. Interestingly, increasing the number of Few-Shot examples provides marginal gains in terms of generability and duplication rate. However, there is a trend in which duplication and equivalent rates increase slightly as the number of Few-Shot examples grows, suggesting that an excessive number of examples limits the LLM's ability in producing diverse and creative mutants.

Overall, the results clearly suggest that LLMs benefit from few-shot examples. However, these findings indicate that although incorporating Few-Shot examples benefits LLMs, carefully selecting the number of examples is important to balance the generation of valid, executable mutants with the generation of diverse and high-quality mutants.

\begin{custombox}{Answer to RQ3}
Prompt strategy has shown to substantially impact on LLM performance in mutant generation. Few-Shot prompts consistently outperform Zero-Shot prompts in both Global and Local mutation strategies. However, too many examples may cause overfitting, so the number of few-shot examples must be balanced.
\end{custombox}

\subsection{$RQ_4$ - Mutation Diversity}\label{Results:RQ4}

To assess the impact of LLMs' temperature parameter on generating mutants, we varied the temperature parameter and compared results across 6 values. Temperature controls the randomness of the model’s sampling process; that is, low values make the model more deterministic, while high values encourage more exploratory and diverse outputs. Usually, the default value of the models ranges from 0.6 to 0.8, with 0.7 being the most common. The results in Table~\ref{tab:RQ4} reveal a clear pattern. At low temperatures (0.2 and 0.4), the models produced the most reliable mutants. These configurations achieved the highest generability (0.933–0.935) and compilability rates (0.736), indicating that the models can consistently generate syntactically correct and executable mutants. These values also showed to be the best at Equivalent Rate, up to 0.123, and mutant quality, up to 0.606, suggesting that deterministic sampling of the LLMs overall favors meaningful and syntactically correct mutants.

\begin{table}[ht!]
\centering
\caption{$RQ_4$ - Comparison of the temperature value impact in the generation of mutants}
\label{tab:RQ4}
\resizebox{0.475\textwidth}{!}{
\begin{tabular}{lrrrrrr}
\toprule
 & \multicolumn{6}{c}{\textbf{Temperature Values}} \\ \cmidrule{2-7}
 & \textbf{0.2} & \textbf{0.4} & \textbf{0.6} & \textbf{0.7} & \textbf{0.8} & \textbf{1} \\ \cmidrule{1-7}
\textbf{Mutant Count} & \textbf{0.975} & \textbf{0.975} & 0.960 & 0.962 & 0.956 & 0.952 \\
\textbf{Generability rate} & 0.935 & 0.933 & \textbf{0.936} & 0.931 & 0.926 & 0.916 \\
\textbf{Compilability rate} & \textbf{0.736} & \textbf{0.736} & 0.731 & 0.722 & 0.721 & 0.700 \\
\textbf{Duplication Rate} & 0.263 & 0.274 & 0.267 & 0.269 & 0.275 & \textbf{0.255} \\
\textbf{Equivalent Rate} & \textbf{0.123} & 0.127 & \textbf{0.123} & 0.128 & 0.143 & 0.137 \\
\textbf{Mutant Quality} & \textbf{0.606} & 0.598 & 0.593 & 0.597 & 0.589 & 0.586 \\
\bottomrule
\end{tabular}}
\end{table}

As the temperature increases, the model becomes more explorato-ry. Higher temperatures (0.8 and 1.0) reduced duplication rate, down to 0.255, showing that randomness helps the models escape repetitive patterns and generate a wider variety of mutants. However, the difference compared to the lower temperature values is marginal. However, this increase in diversity is not purely beneficial as the equivalent mutant rate also increases (i.e., up to 0.143 at a temperature value of 0.8), meaning that some of these diverse mutants are semantically identical to the original model, making them redundant. Additionally, higher temperatures lead to lower generability and compilability, down to 0.916 and 0.700 at a temperature value of 1.0, as the increased randomness introduces more syntactic and semantic errors. Consequently, mutant quality declines at high temperatures. Mid-range temperatures (0.6–0.7) showed slightly lower compilability and mutant quality than the lowest temperatures, but the differences are minimal. Notably, the highest generability rate was observed at 0.6. While mid-range temperatures do not fully match the reliability of the lowest temperatures, they still offer valuable performance.

\begin{custombox}{Answer to RQ4}
Temperature strongly affects mutant generation. Low temperatures produced the most reliable and high-quality mutants, while mid-range temperatures showed similar effectiveness. Overall, despite lower temperatures showing the best performance, default values also yield valuable results.
\end{custombox}

\subsection{$RQ_5$ - Generation Errors}
To understand why some LLM-generated mutants fail, we analyzed a subset of errors occurring during the generation and compilation of mutated models. We found that non-generable mutants primarily result from structural inconsistencies in the proposed mutations. The most frequent issue was invalid references to model elements, where the LLM specified a state or transition ID to mutate that does not exist in the original model. Similarly, some mutations assigned transitions to non-existent source or destination states or junctions, making it impossible to construct a valid Stateflow model.

Non-compilable mutants generally stem from semantic or syntactic issues introduced by the LLM. A common issue we detected was the use of undefined or invented variables, which causes the Simulink compiler to fail due to missing declarations. Another common error was that some mutants contained syntactically invalid or incomplete conditions and actions (e.g., malformed Boolean expressions or missing operators). Finally, in a small number of cases, transitions lost their source or destination during mutation (i.e., the mutation did not introduce any state or junction ID), leading to structurally invalid models that could not be executed. These four error types highlight the need for structural and syntactic validation when using LLMs to generate mutated code, or in this case, Simulink-Stateflow models.

\begin{custombox}{Answer to RQ5}
Generation and compilation errors arise mainly from invalid references to model elements, incorrect transition sources or destinations, invented variables, and malformed logic. These failures underscore the importance of validation mechanisms in LLM-based mutant generation.
\end{custombox}

% \begin{figure}[ht!]
%     \centering
%     \includegraphics[width=1\linewidth, trim = 0 0 0 0]{Figures/RQ5_Forge.pdf}
%     \caption{$RQ_5$ - Hallucination Rate of the evaluated LLMs \pablo{If space needed I can reduce this image's size}}
%     \label{fig:architecture}
% \end{figure}

\section{Threats to Validity}\label{sec:threats}

Our evaluation, like any empirical study, is subject to several validity threats. For \textbf{external validity}, we relied on four Simulink Stateflow models, which may appear limited when compared to traditional state-of-the-art mutant generation studies. However, within the CPS domain, access to a large number of diverse Stateflow models is cumbersome. Therefore, we selected models with varied characteristics, including differences in the number of inputs and outputs, internal structure, and complexity, ensuring that our evaluation captures a representative range of diverse Stateflow models. Another \textbf{external validity threat} is the LLMs evaluated in our study. To mitigate this limitation, we evaluated eight state-of-the-art LLMs, encompassing both open-source and closed-source models with diverse architectures, which strengthens the generalizability of our findings. We also generated a large number of mutants, which also mitigates this threat.

For \textbf{internal validity}, LLM internal configuration parameters such as temperature could influence the performance of LLMs. To mitigate this, we used each model’s default settings and systematically varied the temperature across six values, spanning the full operational range. This allowed us to evaluate the effect of the temperature and ensure a fair comparison across models. In addition, the different prompt settings in our experiments may also threaten the validity of the results. To address this, we explored the impact of prompt strategy and prompt style, including the number of few-shot examples as discussed in Sections~\ref{Results:RQ2} and~\ref{Results:RQ3}.

\textbf{Conclusion validity} is related to the inherent randomness of LLM-generated outputs. We mitigated this by generating a large number of mutants (i.e., 1,200 for each combination of LLM and Stateflow model), resulting in a total of 38,400 mutants. This extensive generation reduces the effect of randomness and strengthens the robustness of our conclusions. 

Finally, \textbf{construct validity} concerns whether the used metrics ensure a meaningful evaluation. To do so, we used eight commonly used metrics from prior mutant generation or mutation testing studies~\cite{wang2024comprehensive, estero2015quality}. These metrics collectively assess syntactic correctness, redundancy, and overall effectiveness, providing a view of LLM's capabilities in effectively and efficiently generating mutants.

\section{Related Work}\label{sec:RelatedWork}

\subsection{AI-based Mutant Generation}
Mutation analysis~\cite{budd1980mutation}, a well-established technique for assessing test quality, has traditionally relied on rule-based mutation operators~\cite{jia2010analysis, just2014major}, such as replacing arithmetic or logical operators from the code. While effective, these approaches are often limited in both expressiveness and realism. To address this limitation, recent studies have explored the potential of LLMs and other AI techniques to improve mutation analysis~\cite{wang2024comprehensive}. For example, $\mu$Bert~\cite{degiovanni2022mu, khanfir2023efficient} adopts a masked-language modeling strategy, making a single token at a time and predicting its replacement. In contrast, LLMorpheus~\cite{tip2025llmorpheusmutationtestingusing} introduces placeholders for abstract syntax tree (AST) nodes and employs prompt-based queries to generate the mutations, enabling more substantial changes to complex expressions. While $\mu$Bert does not guide the masked token, LLMorpheus leverages prompts to guide the generation, generating more diverse and semantically meaningful mutants. 

Building on this direction, Wang et al.~\cite{wang2024exploratory} conducted an exploratory study on LLM-based mutant generation at the method level, moving beyond the AST-node focus of LLMorpheus. In their study, Wang et al. highlight how prompt engineering (temperature, context, guidance) affects the quality of generated mutants. In addition, they demonstrated the effectiveness of LLMs at generating mutants. Similar to our approach, Wang et al. guide the LLM by different prompt templates. Similarly, Brownlee et al.~\cite{brownlee2025large} also demonstrated that carefully designed prompts can improve the diversity of generated solutions since they explored LLMs as mutation operators within genetic and search-based testing. All these studies suggest that LLMs can broaden the scope of mutation testing by generating more realistic and diverse mutants. However, these works primarily target traditional source code, whereas our research focuses on Simulink Stateflow models.

\subsection{Mutant Generation in Simulink-Stateflow Models}
While all the aforementioned approaches~\cite{wang2024exploratory,brownlee2025large,tip2025llmorpheusmutationtestingusing,degiovanni2022mu,khanfir2023efficient,wang2024comprehensive} focus on source code, other studies have investigated mutant generation for Simulink models. Early tools such as SIMULTATE~\cite{pill2016simultate}, provided a user-friendly interface for generating mutants in Simulink models, although the need for manual configuration limited their scalability. The Fault Injection and Mutation tool (FIM)~\cite{bartocci2022fim} represented an important step toward automation, supporting single- and multi-fault injection with a wide range of fault types and mutation operators. However, FIM currently lacks support for generating mutants for Stateflow models. In parallel, the MUT4SLX framework~\cite{ceylan2023mut4slx} has established itself as an open and extensible platform for mutation testing in Simulink. While its initial focus was on block-level mutation, it has recently been extended with operators for Stateflow models~\cite{nuyens2024mut4slx}. Despite these advantages, the operators remain rule-based and manually defined.

BERTiMuS~\cite{zhang2025simulink} used a pre-trained language model (i.e., CodeBERT) to mutate Simulink models by masking tokens and generating mutations, effectively replicating and extending known mutation patterns but with limited Stateflow support. In contrast, our approach leverages prompt-based generative LLMs to directly mutate Stateflow diagrams, including states, transitions, guards, and events, enabling richer and more realistic faults that go beyond token-level changes. This makes our work the first systematic exploration of LLM-driven mutant generation in this domain.

\section{Conclusion}\label{sec:Conclusions}

LLM-based mutant generation offers a promising new direction for mutation analysis for Simulink-Stateflow models, which are widely used in Cyber-Physical Systems (CPS) development. However,to the best of our knowledge their effectiveness across mutation strategies, prompting styles, and temperature settings had not been thoroughly studied in this domain. We present an automated pipeline that transforms Stateflow models into JSON and uses LLMs to generate mutants. In our evaluation, we evaluated eight state-of-the-art models with varying characteristics across eight well-known metrics in mutation analysis. In a large-scale study of 38,400 mutants across four Smimulink-Stateflow models, LLMs generated mutants up to 13x faster than the baseline, with fewer duplicates and equivalents and higher overall quality. We reveal that global mutation with few-shot prompting achieved the highest generability and compilability, while local mutation improved quality but increased duplicate and equivalent rates. Moreover, temperature strongly influenced performance, with low–medium values being most effective. Finally, Error analysis showed four main failure causes: 1) invalid references, 2) incorrect transitions, 3) undefined variables, and 4) syntax errors, highlighting the need for structural and semantic validation. Long story short, LLMs showed to be effective and efficient at Simulink-Stateflow mutant generation.

\section*{Acknowledgments} 
Pablo Valle and Aitor Arrieta are part of the Software and Systems Engineering research group of Mondragon Unibertsitatea (IT1519-22), supported by the Department of Education, Universities and Research of the Basque Country. Pablo Valle is supported by the Pre-doctoral Program for the Formation of Non-Doctoral Research Staff of the Education Department of the Basque Government (Grant n. PRE\_2025\_2\_0252). Shaukat Ali is supported by the Co-tester project (No. 314544) funded by the Research Council of Norway.

\bibliographystyle{ACM-Reference-Format}
\bibliography{bibliografia}

\end{document}